\documentclass[traditabstract]{aa}

\usepackage{graphicx}
\usepackage{txfonts}
\usepackage{natbib}
\usepackage{array}
\usepackage{xspace}
\usepackage{lscape}
\usepackage{url}
\usepackage{arydshln}
\usepackage{caption}
\usepackage{subcaption}

\newcommand{\xabs}{{\it xabs}\xspace}
\newcommand{\hot}{{\it hot}\xspace}
\newcommand{\DELTA}{{\it delta}\xspace}
\newcommand{\amol}{{\it amol}\xspace}
\newcommand{\vgau}{{\it vgau}\xspace}
\newcommand{\xmm}{{\it XMM-Newton}\xspace}

\newcommand{\chandra}{{\it Chandra}\xspace}

\begin{document}

\title{Probing the photoionised outflow in the NLS1 Ark 564:\\ An XMM-Newton view}

\author{Shourya Khanna \inst{1} 
\and Jelle. S. Kaastra \inst{1,2, 3}
\and Missagh Mehdipour \inst{2}}

\institute{Leiden Observatory, Leiden University, PO Box 9513, 2300 RA Leiden, The Netherlands \\{\email{khanna@strw.leidenuniv.nl}} 
\and SRON Netherlands Institute for Space Research, Sorbonnelaan 2, 3584 CA Utrecht, the Netherlands
\and Department of Physics and Astronomy, Universiteit Utrecht, PO Box 80000, 3508 TA Utrecht, The Netherlands}

\date{Received 24 July 2015; Accepted 9 November 2015}
\abstract
{We present a detailed analysis of \xmm X-ray observations of the Narrow line Seyfert-1 (NLS1) galaxy Ark 564 taken between 2000 and 2011. High-resolution X-ray spectroscopy is carried out on the resultant high signal-to-noise stacked spectrum. We find three separate photoionised warm absorbers outflowing at velocities unusually lower than typical NLS1s. Using recombination timescale estimates, improved constraints on the location of these clouds show they could be located beyond 4 pc from the central source. Our estimates of the outflow kinetics suggest that the AGN in Ark 564 is unlikely to affect the host galaxy in its current state but over typical lifetime of 10$^{7}$ years the ISM could be affected. The individual observations used here suggest the luminosity varies over weekly timescales and in addition we find evidence of gas response to changes in the ionising radiation.}
\authorrunning{S. Khanna et al.}
\titlerunning{The photoionised outflow in Arakelian 564}
\maketitle
\section{Introduction}
Active Galactic Nuclei are among the brightest objects in the sky, spanning across 9 orders of magnitude in luminosity \citep{2011ApJS..196....2S}. The engine of this energy output is thought to be accretion onto a supermassive black hole (SMBH) at the centre of the host galaxy resulting in typical luminosities of $\sim$ 10$^{42}$ erg s$^{-1}$ (Seyferts) to 10$^{46}$ erg s$^{-1}$ (Quasars). In addition to a bright continuum, a plethora of absorption and emission features in the X-ray band are also a common feature and in several sources the observed spectral lines are blue-shifted by several 1000 km s$^{-1}$, indicating the presence of outflowing gas in the line of sight. One suggested explanation is that these outflows are produced by irradiation of the dusty gas torus structure \citep{1993ARA&A..31..473A}, which surrounds the SMBH and accretion disk \citep{2001ApJ...561..684K}. 

High resolution X-ray spectroscopy can be performed on nearby sources and this allows us to characterize both the environment as well as the energetics associated with ionised outflows from AGN. This is important in the context of feedback to the host galaxy, which is considered a solution to the galaxy luminosity function bright end problem \citep{2012RAA....12..917S}. But since  only about 10\%\ of AGN are radio-loud \citep{2015JApA..tmp...25C} it seems natural to consider kinetics of the radio-quiet sources such as Seyferts, which are a sub-class of AGN found in the nearby universe. While in Seyfert 2 galaxies both the disk and black hole are obscured by dust (side-on view), in Seyfert 1's both these features are visible (top-down view) making them ideal candidates for studying outflows. In particular NLS1s show broad absorption troughs in the optical and X-ray and also exhibit strong Fe (II) lines \citep{1985ApJ...297..166O} allowing regions close to the central source to be studied. In many low-luminosity Seyfert galaxies partially ionised material or `warm-absorbers' gives rise to absorption features at energies around 1 keV. In the last two decades surveys have found warm-absorption in about 50\% of nearby Seyferts \citep{1994MNRAS.268..405N,2005A&A...431..111B}. In recent years warm-absorbers have also been detected in a few Radio galaxies although the jet still remains the dominant feedback mechanisms in such AGNs \citep{2012MNRAS.419..321T}.

In this paper we attempt to characterise the physical structure of the NLS1 AGN Arakelian 564 (hereafter Ark 564) by making use of all \xmm data on the source so far. Ark 564 is located at a redshift z = 0.02468 and is among the brightest sources in the nearby universe with a 2--10 keV flux f$_{2-10} \sim$ 10$^{-11}$ erg s$^{-1}$cm$^{-2}$. We perform spectral analysis on the stacked spectrum in order to determine the ionisation and dynamical structure of the outflow. Ark 564 like several other NLS1s is already known to be a highly variable source best illustrated during the month long ASCA observation by \citet{2002A&A...391..875G}, when it displayed non-linear behaviour. Given the timespan of our data we will look for long-term variability of the source and the resulting gas response.

Radio-quiet quasars are scaled up versions of Seyferts \citep{2006A&A...455..161L} found at cosmological distances (z > 0.1). Estimating the mass loss in nearby AGN can provide order-of-magnitude estimates of the impact of outflows on host galaxy star-formation \citep{2007ApJ...659.1022K}. We will derive similar quantities and compare with other well known sources.
\section{Observations and data reduction}
\label{data reduction}
The data were extracted using the XMM-Newton SAS software v13.5. Fluxed spectra with 0.01~\AA\ wide bins were created for each RGS detector and each spectral order, and for each individiual observation. Following the methods described by \citet{2011A&A...534A..37K}, the four spectra were combined into one spectrum using the RGS\_fluxcombine program, an auxiliary program of the {\sc spex} software \citep{1996uxsa.conf..411K} version 2.06.00\footnote{http://www.sron.nl/spex}. The same program is used to produce the stacked total spectrum of Ark 564, and the optimised response matrices were produced using the RGS\_fmat program. The details of the 13 \xmm RGS observations are provided in Table \ref{obs_table}. The average exposure time is about 50 ks. In order to obtain the best signal-to-noise ratio, the individual observations have been stacked together to produce a time-averaged spectrum  discussed in Sect. \ref{Spectral fitting}.
%
\begin{table*}[!tbp]
\begin{minipage}[t]{\hsize}
\setlength{\extrarowheight}{3pt}
\caption{Details of the \xmm (RGS) observations of Ark 564. Hereafter, in this paper the sources will be called by the observation tags shown here.}
\label{obs_table}
\centering
\renewcommand{\footnoterule}{}
\begin{tabular}{c c c c c}
\hline\hline
& & \multicolumn{2}{c}{Start time (UTC)} & RGS X-ray exposure \\
Observation tag & Observation ID &yyyy-mm-dd & hh:mm:ss &  (s) \\ 
\hline
01 & 0006810101 &2000-06-17 &03:48:06 &   16962   \\
02 & 0006810201 &2000-06-17 &08:46:32 &   22140   \\
03 & 0006810301 &2001-06-09 &05:20:12 &    6944   \\
04 & 0006810401 &2001-06-09 &07:25:45 &   15656   \\
05 & 0206400101 &2005-01-05 &19:15:22 &   100630  \\
06 & 0670130201 &2011-05-24 &05:56:05 &   59316   \\
07 & 0670130301 &2011-05-30 &15:08:42 &   55753   \\
08 & 0670130401 &2011-06-05 &22:56:32 &   62445   \\
09 & 0670130501 &2011-06-11 &16:59:18 &   67098   \\
10 & 0670130601 &2011-06-17 &04:16:50 &   60742   \\
11 & 0670130701 &2011-06-25 &23:16:43 &   54611   \\
12 & 0670130801 &2011-06-29 &06:48:05 &   58056   \\
13 & 0670130901 &2011-07-01 &06:45:51 &   55747   \\
\hline
\end{tabular}
\end{minipage}
\end{table*}

\section{Spectral Analysis}
\label{spectral analysis}
\subsection{Source Spectral Energy Distribution}
\label{SED_preparation}
%
\begin{figure}
\resizebox{8.5cm}{!}{\includegraphics[angle=0]{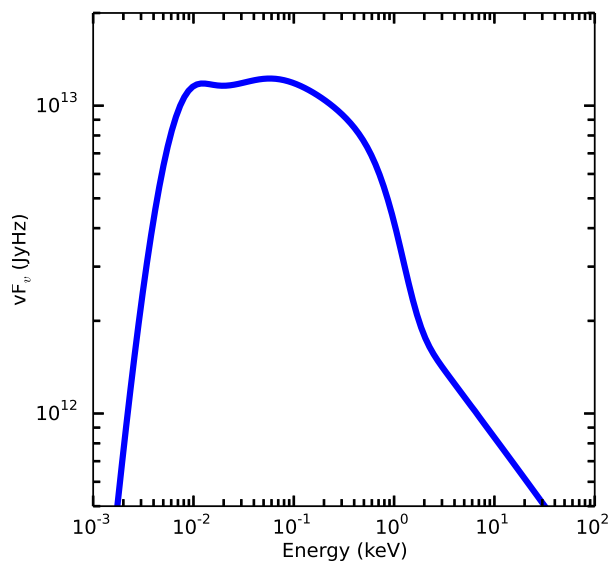}}
\caption{Spectral Energy distribution of Ark 564 derived using the EPIC-pn and OM data of 2011 \xmm observations.}
\label{ark564 SED 2011}
\end{figure}
For the purpose of photoionisation modelling of the outflows, we determined the broadband spectral energy distribution (SED) of the ionising source in Ark 564 (Fig. \ref{ark564 SED 2011}). The UV and X-ray parts of the SED were constructed from the \xmm OM photometric filters (UVW1, UVM2 and UVW2) and the EPIC-pn spectrum (0.3--10 keV).
 
In order to expand the energy coverage of the SED beyond that of \xmm into lower and higher energies, we used archival flux measurements from the NASA/IPAC Extragalactic Database (NED). We extracted infrared flux measurements at 25, 60 and 100 micron from IRAS and a radio flux measurement at 2380 MHz from Arecibo. We also included a Suzaku flux measurement given over the 10--50 keV band. As the source becomes too faint at hard X-rays, we applied a cut-off at 400 keV to our continuum model, which is a typical cut-off value found in bright AGN eg. NGC 5548 \citep{2015A&A...577A..38U}.
 
We determined the SED by applying a broadband AGN continuum model to the above data. We used the broadband continuum model discussed in Mehdipour et al. (2011, 2015), which is based on Compton up-scattering of the IR/optical/UV disk photons to X-ray energies in warm and hot coronae. The use of this model enables us to establish the continuum from infrared to X-ray energies, including the EUV part, which is important for photoionisation calculations.
 
In addition to correction for interstellar X-ray absorption and reddening in the Galaxy, we also corrected for reddening in the host galaxy of Ark 564. There are reports of substantial internal reddening in Ark 564. Thus, we adopted the colour excess E(B -- V) = 0.14 reported by \citet{2002ApJ...566..187C} to correct for host galaxy reddening in Ark 564, assuming the same dust extinction law as the Galaxy. In our galaxy, the colour excess is smaller at E(B -- V) = 0.03 \citep{1998ApJ...500..525S}. To correct the data, we applied the reddening curve of \citet{1989ApJ...345..245C}, including the update for near-UV given by \citet{1994ApJ...422..158O}.

All spectra were modelled using the {\sc spex} code, adopting a $\Lambda$-cosmology with $\Omega_{\mathrm{\Lambda}}$=0.7, $\Omega_{\mathrm{M}}$=0.3 and $H_{\mathrm{0}}$ = 70 km s$^{-1}$ Mpc$^{-1}$. For the spectral analysis we use $\chi^{2}$ statistics and the optimum data bin size was determined using Shannon binning. Adopting this the data were binned with a redistribution function FWHM of $\bigtriangleup$E/3. The RGS resolution is 70 m$\AA$ (Shannon FWHM $\sim$ 50 points/$\AA$) while our stacked spectrum has 100 points/$\AA$, thus we have binned our data by a factor of 2.\\

\subsection{Continuum Fitting}
\label{Spectral fitting}

In previous studies, the continuum has usually been modelled with a steep power law of photon index$\sim$2.5 and soft-blackbody components \citep{2004MNRAS.347..854V}. But others have also shown that the soft-part is not very well modelled with a simple black-body, there have been relativistic disk profiles as well as multiple black-body components in one fit \citep{2007A&A...461..931P}. Taking a cautious approach we use spline interpolation to fit the basic continuum. A grid of 18 points equally spaced by 2 $\AA$ was used over the range 6-40 $\AA$ with an extended boundary to reduce errors due to sharp discontinuity at the edge of the RGS band.

\begin{figure}
\centering
\resizebox{7cm}{!}{\includegraphics[angle=0]{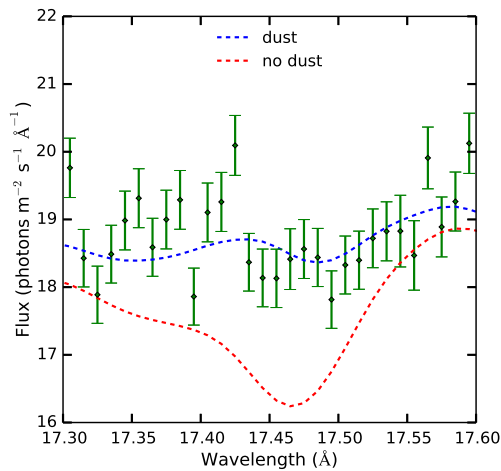}}
\caption{Weakness of galactic neutral iron in gas form (red curve) in the line of sight to Ark 564 predicted by N$_{H}$ = 5.34 $\times$ 10$^{20}$ cm$^{-2}$ \citep{2005A&A...440..775K}, suggests presence of Fe in dust form (blue curve) in our own galaxy.}
\label{fe_dust_17}
\end{figure}

For the Galactic interstellar X-ray absorption we used the {\sc spex} \hot model with $N_{\mathrm{H}}$ = 5.34 $\times$ 10$^{20}$ cm$^{-2}$ in our line of sight towards Ark 564 \citep{2005A&A...440..775K}. This is assumed to be a cold neutral absorber ($kT\sim$0.5 eV) and clearly fits the O I Galactic oxygen line at 23.5 $\AA$ in the RGS spectrum. This was also identified earlier by \citet{2013A&A...551A..95R} and \citet{2008A&A...490..103S}. However, the \hot model also predicts absorption around 17.47 $\AA$ corresponding to neutral iron. The data do not show a strong signature here and it is likely due to Galactic dust. By freeing the atomic Fe abundance in the \hot model and including iron in dust form (using the \amol model in {\sc SPEX} and component Fe$_{2}$O$_{3}$) we get a much better fit (see the blue line in Fig. \ref{fe_dust_17}) with an Fe$_2$O$_3$ column of $N_{\mathrm{H}}$ = 3.36 $\times$10$^{15}$ cm$^{-2}$. This is consistent with other high-resolution X-ray spectra through the Galactic ISM, e.g. \citet{2013A&A...551A..25P}.
It is not unlikely that the Galactic foreground medium has multiple temperature components and indeed additionally we notice O VII absorption at 21.6 $\AA$ which is well fitted with a second warmer \textit{hot} model with a temperature $\sim$ 0.15 keV and turbulence with $v_{\mathrm{turb}}$ = 36 $\pm$ 20 km s$^{-1}$. This is the 1s$^{2}$ g$^1$S$_0$ - 1s2p $^1$P$_1$ transition \citep{1985A&AS...62..197M}.
%
\subsection{Emission lines}
%
\begin{figure}[!]
\begin{center}
\resizebox{9cm}{!}{\includegraphics[angle=0]{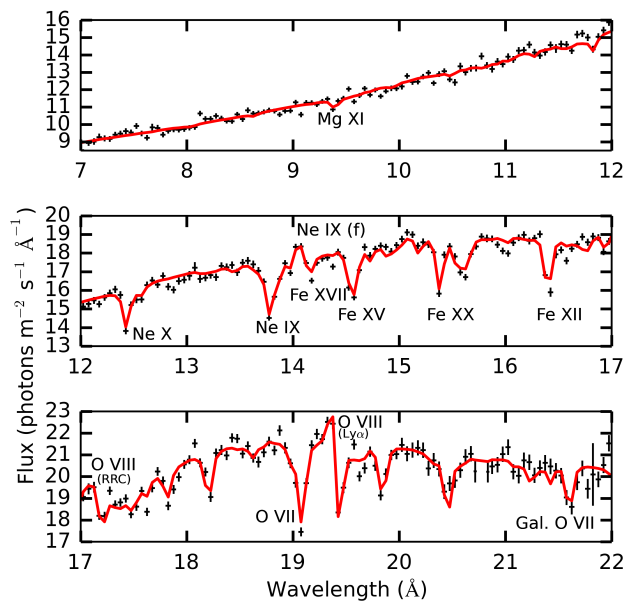}}
\caption{Blow-up of the high resolution RGS spectrum of Ark 564 showing important emission and absorption lines between 7-22 $\AA$ as discussed in the text. The best-fit final full spectrum (7-38 $\AA$) with residuals is shown in Fig. \ref{RGS spectrum of ARK564}.}
\label{spex zoom_7_to_22}
\end{center}
\end{figure}

%
\begin{figure}[!]
\begin{center}
\resizebox{9cm}{!}{\includegraphics[angle=0]{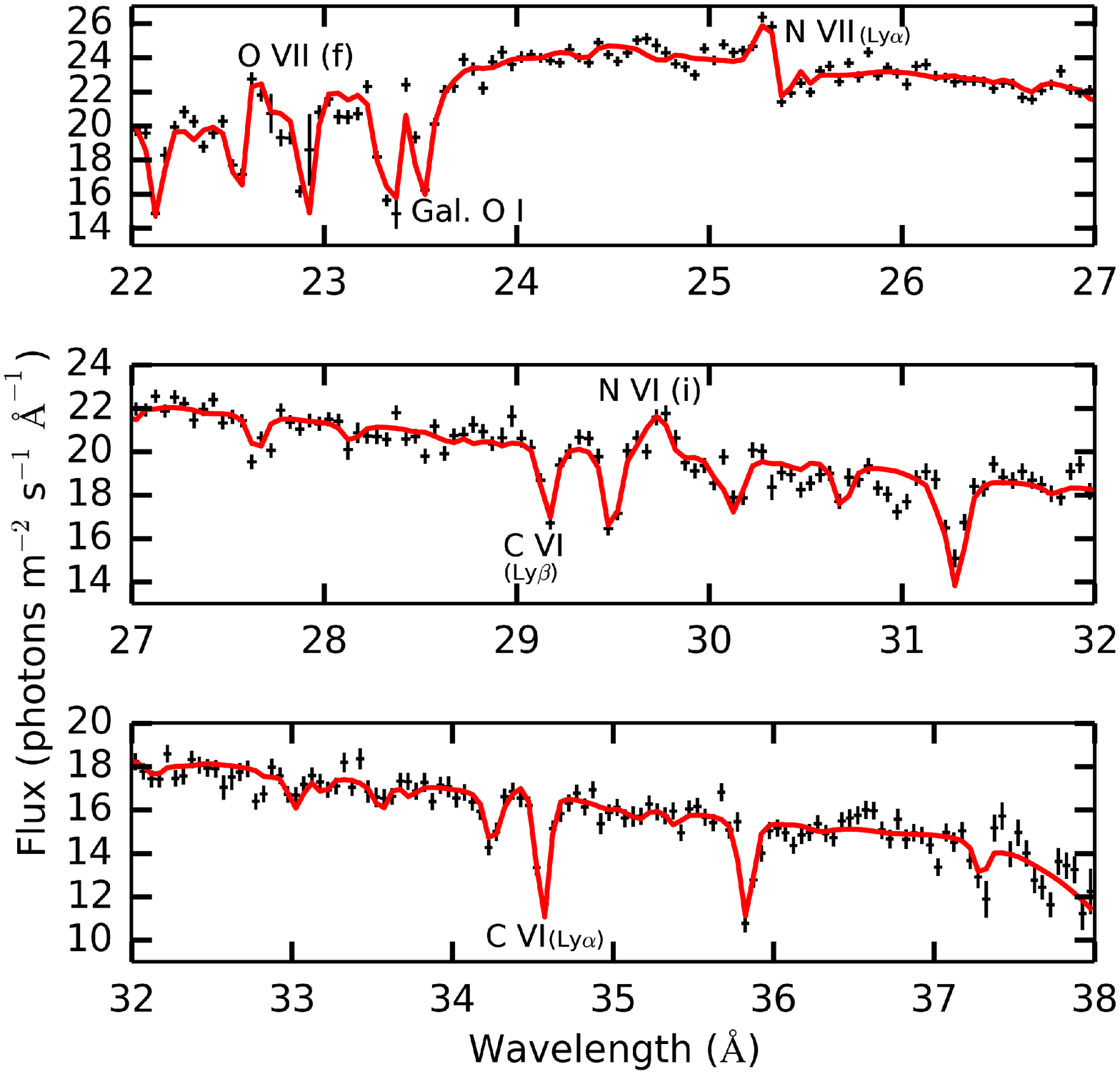}}
\caption{Same as Fig. \ref{spex zoom_7_to_22} but between 22-38 $\AA$.}
\label{spex zoom_22_to_38}
\end{center}
\end{figure}

Figures \ref{spex zoom_7_to_22} and \ref{spex zoom_22_to_38} show zoomed-in sections of the main spectrum, highlighting the emission features. These were fitted with the \DELTA and \vgau models in {\sc spex} which incorporate narrow emission lines with Gaussian broadening. Prominent lines are listed in Table \ref{emission} along with the broadening velocity $\sigma$. 

We infer three distinct zones the emission lines could originate from. The Ly$\alpha$ lines for C, N and O seem to be from a high-velocity/turbulence zone, while the two forbidden lines (O VII and Ne IX) show negligible outflow signature. In addition, the intercombination line of N VI is half as broadened as that of the Ly$\alpha$ lines. \citet{2008A&A...490..103S} suggest that since signature of intercombination lines is an indication of high-density and since the N VI(i) and O VII(f) \&\ Ne IX(f) seem to come from separate components this suggests that the faster component could also be denser. However, we note that for oxygen the derived density estimates from line ratios is likely unreliable due to self-absorption of the resonance and then the intercombination line \citep{2015arXiv150506034M}.
%
\begin{figure}
\resizebox{9cm}{!}{\includegraphics[angle=0]{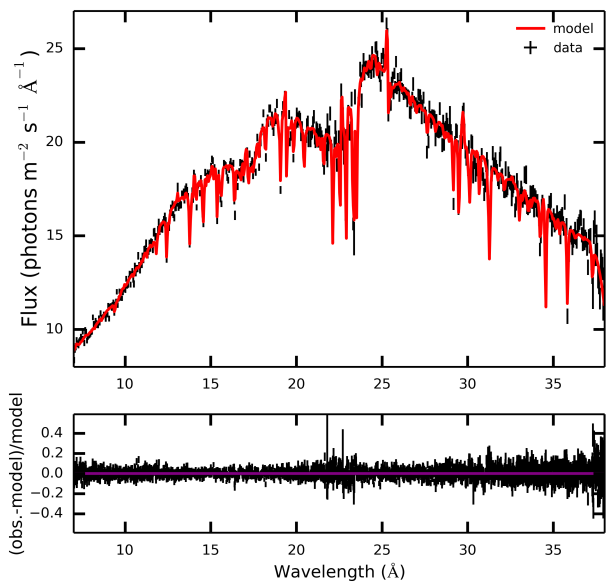}}
\caption{Shown in black is the stacked \xmm RGS high-resolution spectrum produced by combining all Ark 564 RGS observations to date. The final best-fit is shown in red and consists of galactic absorption, broad and narrow Gaussian emission lines, three phases of photoionised warm-absorbers all on a continuum modelled by spline interpolation. The lower panel shows the residuals (black) for this fit.}
\label{RGS spectrum of ARK564}
\end{figure}

The narrow and broad emission lines are generally assumed to arise from a photoionised plasma, where in general the temperature is much smaller than the ionisation potential of the dominant ions in the plasma. In this case, we expect to see Radiative Recombination Continua (RRC). Candidates for RRC in our band include O VII, C V and Ne IX. While we do not detect carbon or neon RRC, there is clearly a feature around 17 $\AA$ which corresponds to the O VIII to O VII recombination. In addition we find a Doppler shift ($z_{\mathrm{Dop}} = \frac{\Delta \lambda}{\lambda}$) at the source for this feature of $z_{\mathrm{Dop}}$= -2.43$\times$10$^{-3}$ corresponding to a blueshift of about 800 km s$^{-1}$.

\subsection{Relativistic Broad lines?}
%
\begin{figure}[!]
\centering
\begin{subfigure}[b]{0.4\textwidth}
\includegraphics[width=\textwidth]{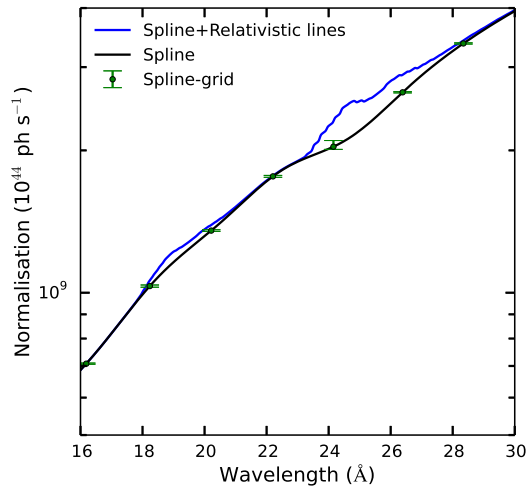}
\end{subfigure}
\caption{The spline continuum with fitting for broad relativistic lines. The error-bars are insignificant as is illustrated in the blow-up of the relevant region: N VII (24.73 $\AA$) and O VIII (18.94 $\AA$) in the laboratory frame.}
        \label{laor significance}
\end{figure}

From the stacked spectrum two extremely broad asymmetric features seem apparent in the region covering 18--30 $\AA$. We explore the possibility of these being broad relativistic emission lines originating from the O VIII and N VII Lyman alpha transitions. Relativistically broadened skewed lines result from a combination of gravitational redshift due to the deep potential around black-holes and relativistic beaming due to gas swirling at relativistic velocities \citep{2002MNRAS.335L...1F}. There have been previous reports of broad Ly$\alpha$ lines for other sources such as MCG-6-30-15 and Mrk 766 \citep{2001A&A...365L.140B}. We checked this broadened feature in the Ark 564 spectrum by modelling with a {\sc laor} profile \citep{1991ApJ...376...90L} in {\sc spex}. The best fit results for an inner radius of 10$\frac{GM}{c^{2}}$, consistent with a non-rotating black-hole, with the disk aligned at about 37$\degr$. The parameters are provided in Table \ref{Stacked best-fit parameters} and including the possible broad relativistic lines at the O VIII and N VII Ly$\alpha$ energies improves $\Delta \chi^{2} \sim$ 500 for 1490 $d.o.f$. Fig. \ref{laor significance} shows the spline continuum with and without the two relativistic lines and though it would appear that the features are statistically significant (given the small error bars) we would like to point out that with a slightly higher, more powerlaw-like spline continuum in this band, the remaining residuals are of the order of a few percent. Recent calibration work by one of us (J.S.K.) shows that the RGS effective area has remaining residuals at a similar level of a few percent on scales of a few $\AA$. We leave it open here whether the relativistic lines are real or serve to compensate for small calibration uncertainties.
\subsection{Photoionisation modelling of absorption clouds}
\label{Absorption clouds}
A preliminary search for absorption features was done using the {\sc slab} model \citep{2002A&A...386..427K} in {\sc spex}, which allows for ionic column densities to be chosen independently. This also provided us with an estimate of the outflow velocity of the absorbing gas. For a more realistic calculation we made use of \xabs components in {\sc spex}. The \xabs model calculates the transmission of a slab of material, where all ionic column densities are linked through the {\sc Cloudy} \citep{1998PASP..110..761F} photoionisation model. The parameters fitted for each \xabs component are the ionisation parameter ($\xi = \frac{L_{\mathrm{ion}}}{n(R) R^{2}}$), turbulence, outflow velocity and the column density with $n(R)$ defined as the hydrogen gas density at $R$ which is the distance between the source and the absorber. We made use of the derived SED of Ark 564 (Sect. \ref{SED_preparation}) in order to calculate the ionisation balance in the {\sc Cloudy} code and obtain $L_{\mathrm{ion}}$ = 6.9 $\times$ 10$^{37}$ W. Here we assume Solar abundances \citep{2009LanB...4B...44L}, however, since AGN are typically known to have high abundances in the elements C, N, O and Fe \citep{2001A&A...374..914K} we also tested for non-solar abundances but did not find strong deviations from solar metal abundance ratios, a similar result as found in the case of Mrk 509 by \citet{2011A&A...534A..42S}. Hence, we have assumed Solar abundances for Ark 564. We find that the best fit is achieved with three separate \xabs components with equal spatial covering factor f$_{\mathrm{cov}}$=1. These vary in ionisation as well as in their outflow signature. Table \ref{absorption phases} ranks the three absorption phases identified by their ionisation parameter ($\xi$) and additionally we note that the highly ionised phase 3 produces most of the Fe and Ne lines. Compared with the outflow velocities for emission lines in Table \ref{emission}, the absorbing gas seems to be present in weakly outflowing regions suggesting that the origin of the absorber and the emission lines is likely to be different. However, the OVII(f) does  have broadening comparable to phases 2 and 3 so there could be a connection between these.\\ 
%
\begin{figure}[!]
\centering
\resizebox{9cm}{!}{\includegraphics[angle=0]{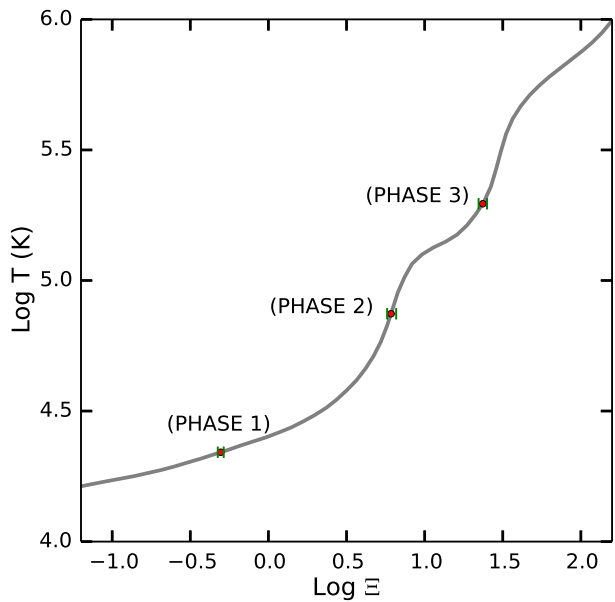}}
\caption{Stability Curve with the positions indicated of the three warm-absorbers from our best fit for Ark 564. The absence of a region with negative slope on the S-curve indicates there are no unstable regions i.e. gas can exist in equilibrium throughout the curve. This is likely due to the steep soft-excess seen in Ark 564 which allows gas to cool efficiently by lowering the Compton temperature. (Marker size exaggerated to show against small error bars in green)}
\label{pressure equilibrium}
\end{figure}
%
\begin{table*}[t]
\begin{minipage}[t]{\hsize}
\setlength{\extrarowheight}{3pt}
\caption{Narrow and Broad non-relativistic emission lines with line-flux as observed}
\label{emission}
\centering
\renewcommand{\footnoterule}{}
\begin{tabular}{cccccc}
\hline \hline
Line & $\lambda$ (observed) & $\lambda_{0}$ (rest) & $\lambda_{0}$ (lab.)& Broadening ($\sigma$) & Line-flux (observed)\\
     &  $\AA$               &  $\AA$ &  $\AA$    & km s$^{-1}$           & 10$^{50}$ ph s$^{-1}$   \\
\hline
O VII (f)    & 22.65$\pm$0.01 & 22.10$\pm$0.01 & 22.10 & 70  $^{+180} _{-70}$  & 0.89 $\pm$0.08 \\[5pt]

Ne IX (f)    & 14.05$\pm$0.01 & 13.71$\pm$0.01 & 13.70 & 0 $\pm$290    & 0.22 $\pm$0.03         \\[5pt]
 
N VI (i)           & 29.76$\pm$0.01 & 29.04$\pm$0.01 & 29.09 & 540 $\pm$150  & 1.00 $\pm$0.16 \\[5pt]

O VIII Ly$\alpha$  & 19.41$\pm$0.06 & 18.94$\pm$0.06 & 18.97 & 770 $\pm$70 & 1.76 $\pm$0.11 \\[5pt]

N VII Ly$\alpha$   & 25.34$\pm$0.02 & 24.73$\pm$0.02 & 24.78 & 1026 $\pm$130 & 2.33 $\pm$0.20 \\[5pt]

C VI Ly$\alpha$    & 34.54$\pm$0.03 & 33.71$\pm$0.03 & 33.74 & 950 $^{+300} _{-470}$ & 1.27 $\pm$0.38 \\[5pt]
\hline
\end{tabular}
\end{minipage}
\end{table*}
%
\begin{table*}[t]
\begin{minipage}[t]{\hsize}
\setlength{\extrarowheight}{3pt}
\caption{Summary of the spectral features (other than narrow emission lines) and best-fit parameters used to model the stacked spectrum of Ark 564 resulting in a final $\chi^{2} / \mathrm{d.o.f.}$= 5178 /3034.}
\label{Stacked best-fit parameters}
\centering
\renewcommand{\footnoterule}{}
\begin{tabular}{cccc}
\hline \hline
Feature & Parameter & Status & Value \\
\hline
Galactic Absorption (cold phase) & & & \\
& nh, Column Density (10$^{24}$ m$^{-2}$) & frozen & 5.34  \\
& Temperature (10$^{-4}$ keV)             & frozen & 5.0    \\
& RMS Turbulence (km s$^{-1}$)            & thawn  & 21 $\pm$ 5  \\
\hline
Galactic Absorption (warm phase) & & & \\
& nh, Column Density (10$^{23}$ m$^{-2}$) & thawn & 2.5 $\pm$ 0.6   \\
& Temperature (keV)                       & thawn & 0.15 $\pm$ 0.01  \\
& RMS Turbulence (km s$^{-1}$)            & thawn & 36 $\pm$ 15 \\
\hline
Radiative Recombination (O VIII $\xrightarrow{}$ O VII) & & & \\
& Emission strength (10$^{64}$ m$^{-3}$) & thawn  & 4400 $\pm$ 400   \\
& Temperature (10$^{-3}$ keV)             & thawn  & 9.5 $\pm$ 0.8   \\
& Doppler shift (10$^{-3}$)               & thawn  & -2.7 $\pm$ 0.6    \\
& Doppler velocity (km s$^{-1}$)          & thawn  & -800 $\pm$ 200  \\
\hline
Broad Relativistic Lines\\
&O VIII Ly$\alpha$ (10$^{50}$ ph s$^{-1}$) & thawn & 4.4 $\pm$ 0.6  \\
& N VII Ly$\alpha$ (10$^{50}$ ph s$^{-1}$)  & thawn & 21.0 $\pm$ 2.0\\
& Inner radius R$_{in}$  (GM/c$^{2}$)       & thawn & 10$^{+2} _{-7}$  \\
& Emissivity index q                        & thawn & 1.83 $\pm$ 0.05\\
& Inclination i                             & thawn & 37$^{\circ} \pm$ 0.5$^{\circ}$  \\
\hline
\end{tabular}
\end{minipage}
\end{table*}
On the basis of the distinct $\xi$ parameters that we find, it would appear that Ark 564 has three separate absorption components. To test this further in Fig. \ref{pressure equilibrium} we plot the pressure form of the ionisation parameter, $\Xi = L_{\mathrm{ion}}/4\pi r^{2} cn_{\mathrm{H}}kT$, against the absorber temperatures ($\approx 10^{5}$K) derived using {\sc spex} with the SED from the 2011 (Fig. \ref{ark564 SED 2011}) observations. In Fig. \ref{pressure equilibrium} regions with positive gradient represent where gas is in photo-ionisation equilibrium and is stable against thermal perturbations while the negative slopes indicate unstable zones. Since at a given distance $r$, $L_{\mathrm{ion}}$ is independent of gas density, $\Xi \propto P_{\mathrm{gas}} ^{-1}$, and thus on such a plot, absorbers with a range of densities and temperatures can line up vertically if in pressure-equilibrium. This is considered to play a role in confining neighbouring outflowing clouds such as in NGC 985 where the absorbers are present in a multi-phase wind \citep{2005ApJ...620..165K}. For Ark 564, it is clear that the three absorbers do not line-up on the stability curve and so are not in pressure-equilibrium. It is interesting to note that the kinetics of phases 1 and 2 are very similar with regards to the outflow velocities as compared to the negligible outflow in phase 3. Additionally, there is overlap in the locations of the absorbers (Sect. \ref{dist_est}) from phases 1 and 2 so perhaps this suggests another mechanism by which these two zones are confined. Distinct ionisation components are also seen in other well studied sources. In NGC 5548 \citet{2005A&A...434..569S} detect up to 5 components not in pressure-equilibrium yet with similar kinetics to each other pointing towards a single confined outflow, quite unlike our situation. Also, Mrk 509 \citep{2011A&A...534A..40E} shows three distinct ionisation zones, although, on the S-curve the outflowing components are in pressure-equilibrium with each other while the third component, which was in fact found to be redshifted was not in equilibrium with its outflowing counterparts. 
%
\begin{table*}[t]
\begin{minipage}[t]{\hsize}
\setlength{\extrarowheight}{3pt}
\caption{The three warm absorber components found in Ark 564. Two of the gas components are outflowing at low velocities while the third phase is consistent with no outflow. The location of each phase has been constrained as discussed in Sect. \ref{dist_est}.}
\label{absorption phases}
\centering
\renewcommand{\footnoterule}{}
\begin{tabular}{cccccccc}
\hline \hline
Phase & Outflow velocity & Turbulence    & Log $\xi$      & N$_{H}$              & Log L$_{abs}$ & C$_{v}$ & Location \\
      & (km s$^{-1}$)    & (km s$^{-1}$) & (10$^{-9}$ Wm) & (10$^{24}$ m$^{-2}$) & (W)           &         & (pc)     \\
\hline
Phase 1 & -48 $\pm$ 15 & 130 $\pm$ 9  & -0.25 $\pm$ 0.02 & 1.56 $\pm$ 0.08 & 37.0 & 0.03  &  80 $\leq$ r $\leq$ 74000 \\        [5pt]
Phase 2 & -53 $\pm$ 17 &  79 $\pm$ 4  &  1.37 $\pm$ 0.03 & 3.18 $\pm$ 0.12 & 36.2 & 0.17  & 7.5 $\leq$ r $\leq$864 \\        [5pt]
Phase 3 &   0 $\pm$ 18 &  87 $\pm$ 5  &  2.38 $\pm$ 0.03 & 7.39 $\pm$ 0.39 & 35.6 & -\footnote{Phase 3 shows negligible outflow which would result in a non-physical covering factor, hence we ignore it in our calculations.} & 4.2 $\leq$ r $\leq$37 \\
\hline
\end{tabular}
\end{minipage}
\end{table*}
For Ark 564 we also note that the S-curve does not have any unstable regions which is likely due to the source having a steep soft X-ray spectrum which removes instabilities by lowering the Compton temperature of the gas, allowing it to cool more efficiently \citep{1983ApJ...266..466G}. Such behaviour has also been noted for other NLS1s such as I ZW 1 \citep{2007MNRAS.378..873C}.

\section{Discussion}
\label{discussion}
\subsection{Comparison with previous spectroscopic studies}

The data sets used in this paper have been studied separately before. Generally the features we report are similar to other works. We obtain similar column density and ionisation parameter to \citet{2013A&A...551A..95R} who observed Ark 564 using Chandra LETGS. The analysis of the line at observed wavelength of 19 $\AA$ suggests this is from phase 2 and we get a slightly higher ionisation parameter of Log $\xi =$1.34 to their Log $\xi =$1.1 but very similar column density and line width. \citet{2008A&A...490..103S} also show this line but do not discuss it. \citet{2015arXiv150201338G} using a different photoionisation code in {\sc xspec} find similar features to us.

Previous works mostly find the best-fit model to be consistent with two separate phases of warm-absorbers. From the Ark 564 2005 observations, \citet{2007A&A...461..931P} find two phases with Log $\xi \sim$1 and 2, similar to the 2002 results from \citet{2004ApJ...603..456M}. \citet{2007ApJ...671.1284D} using both EPIC and RGS data from 2005 report similar numbers and in addition find very high velocity outflows of up to 1000 km s$^{-1}$. \citet{2008A&A...490..103S} performed spectroscopy on the stacked spectrum of data up to 2005 and although they assume continuum parameters as reported by \citet{2007A&A...461..931P}, they found the best-fit model to require three separate phases of absorption and not in pressure-equilibrium similar to the results presented in this paper. 
\subsection{Luminosity and Ionisation variability}
\label{variab_results} 
\begin{figure*}
    \centering
    \begin{subfigure}{0.45\textwidth}
        \resizebox{8cm}{!}{\includegraphics[angle=0]{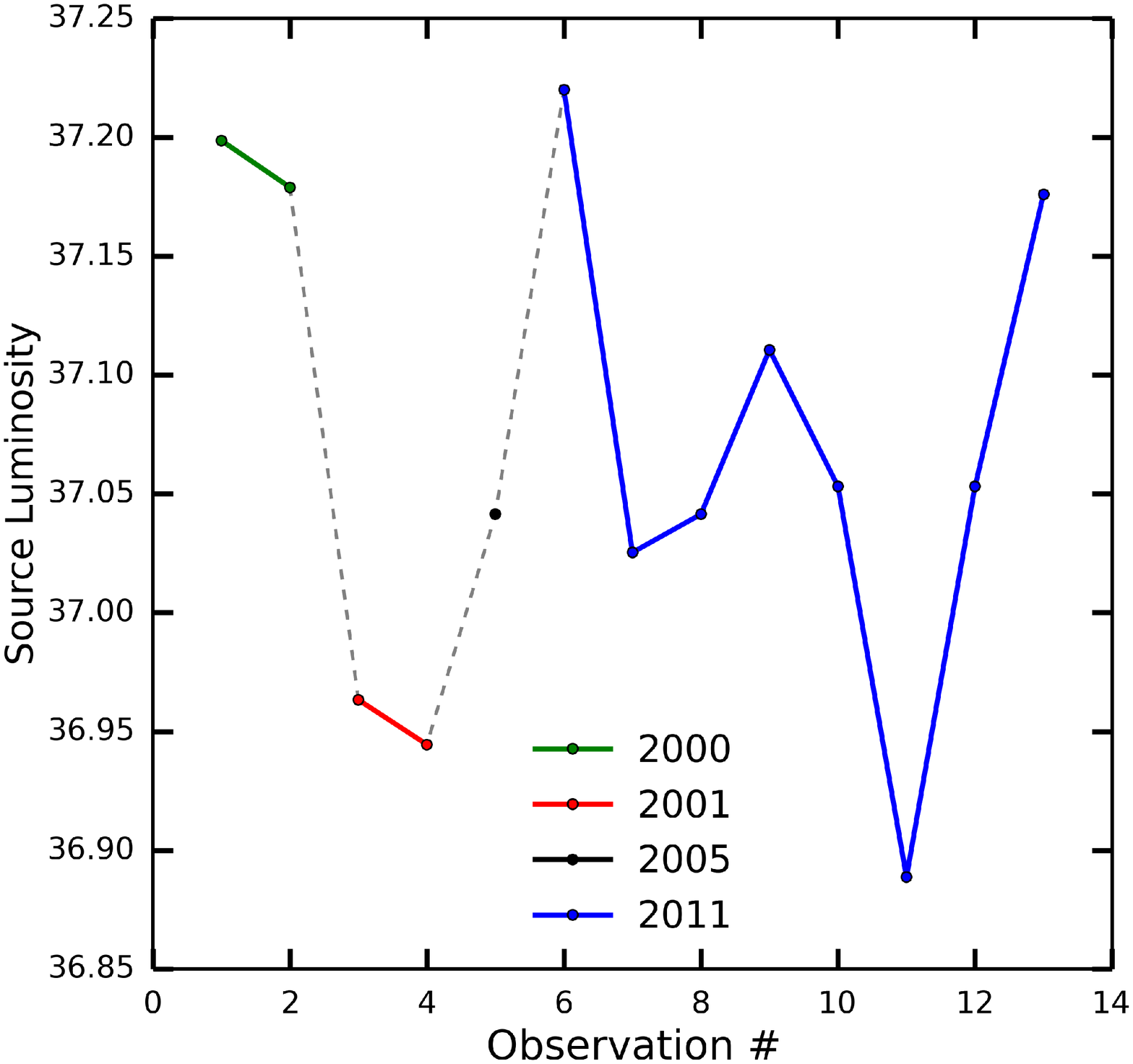}}
        \caption{}
        \label{light curve}
    \end{subfigure}
    \begin{subfigure}{0.45\textwidth}
        \resizebox{8cm}{!}{\includegraphics[angle=0]{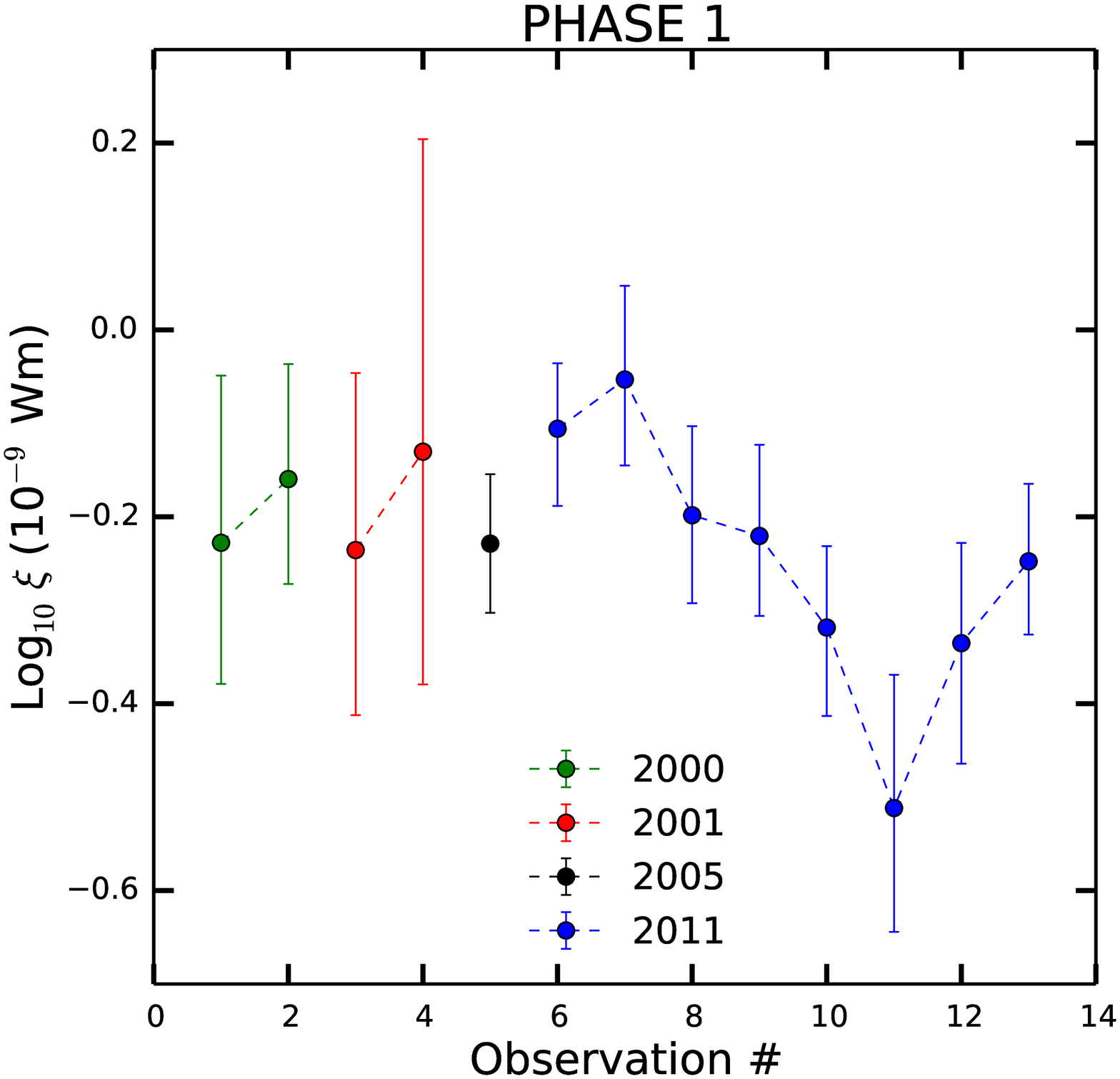}}
        \caption{}
        \label{ionization variability combined ph1}
    \end{subfigure}
    \begin{subfigure}{0.45\textwidth}
        \resizebox{8cm}{!}{\includegraphics[angle=0]{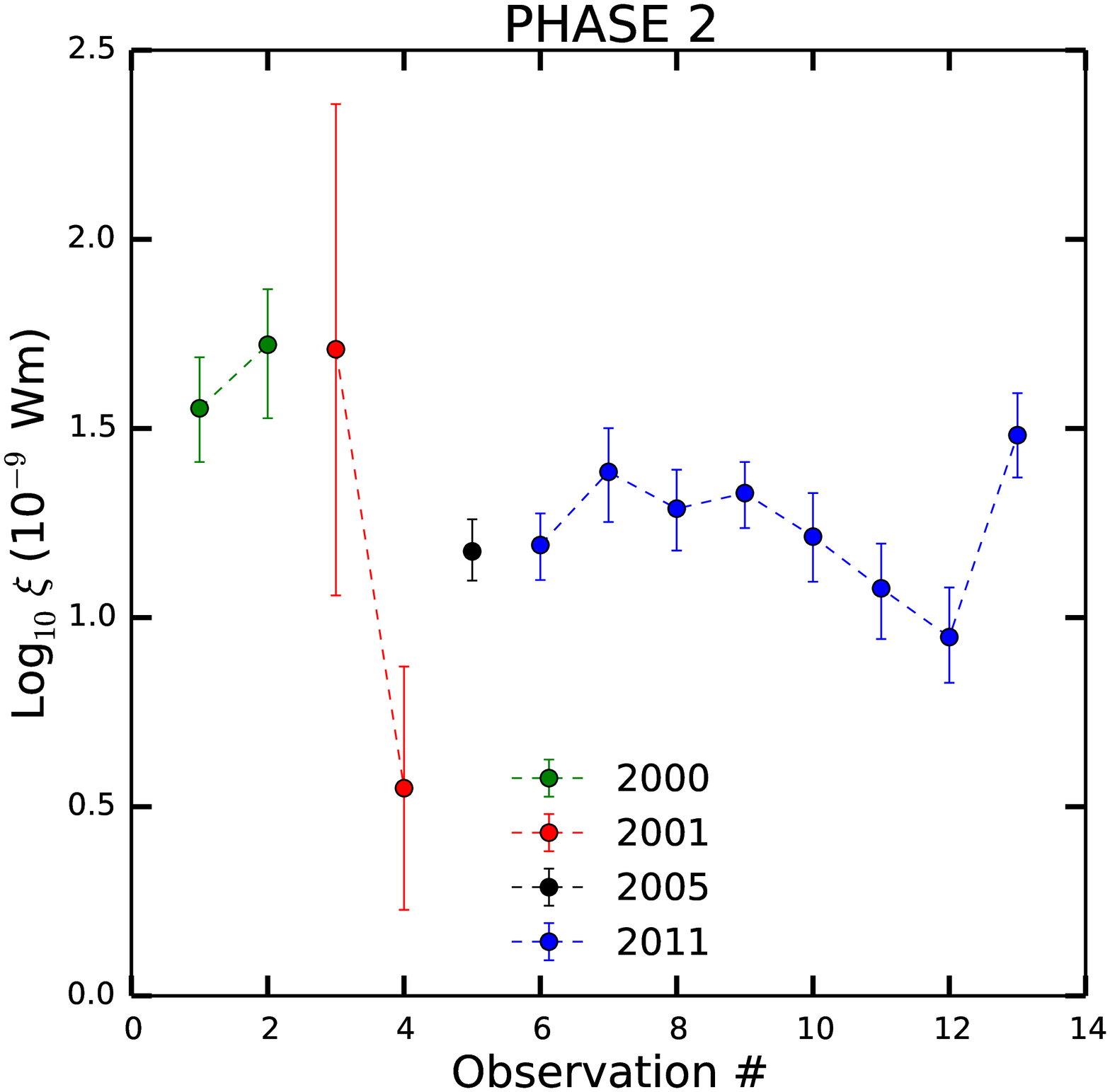}}
        \caption{}
        \label{ionization variability combined ph2}
    \end{subfigure}
    \begin{subfigure}{0.45\textwidth}
        \resizebox{8cm}{!}{\includegraphics[angle=0]{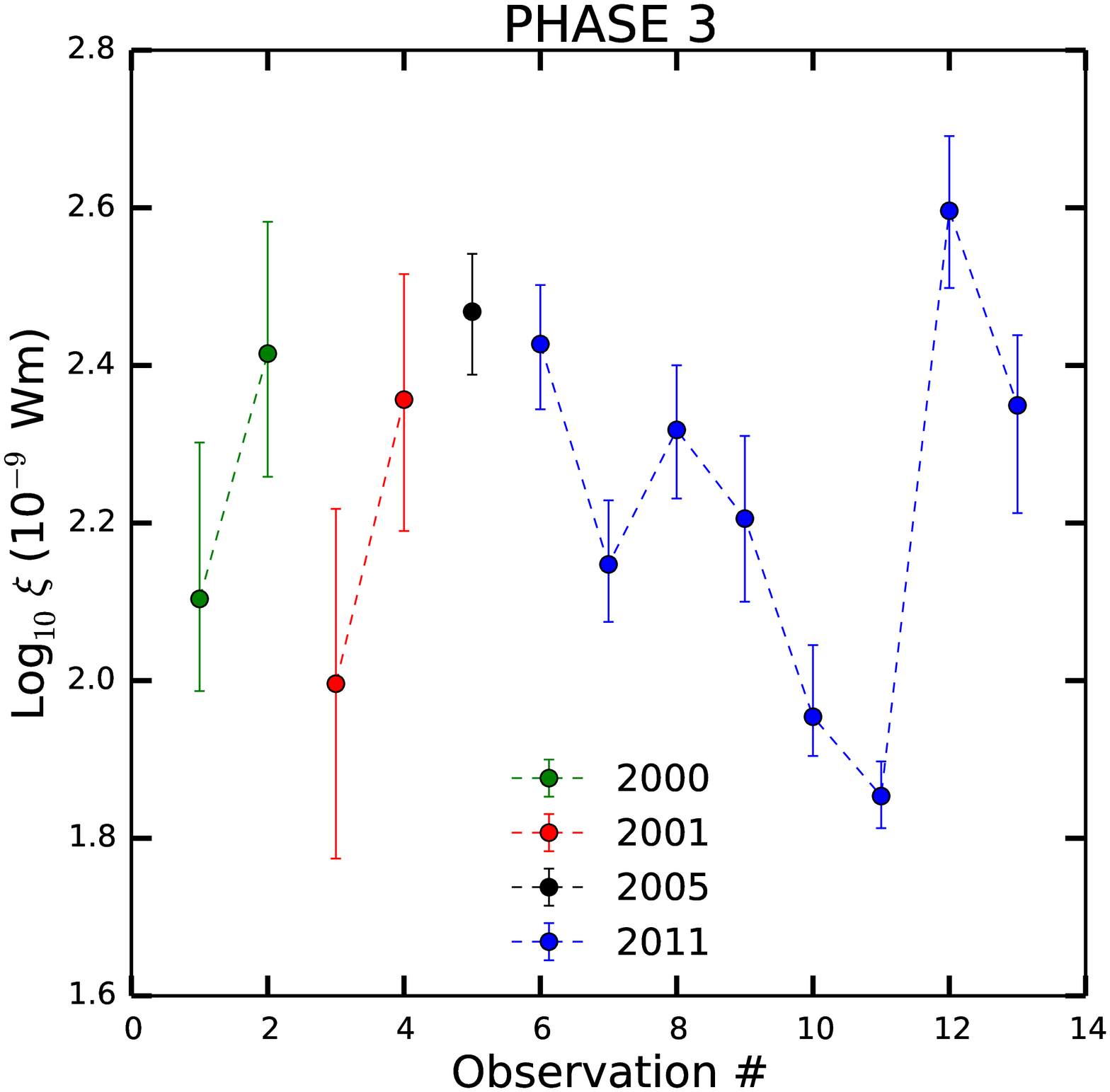}}
        \caption{$\quad$}
        \label{ionization variability combined ph3}
    \end{subfigure} 
    \caption{ (\subref{light curve}) Based on the best-fit parameters from the stacked spectrum we derive luminosities for all the individual observations between 2000-2011. The individual data sets are highlighted and so the combined plot is not evenly spaced in time. Observation \#\ are as in Table \ref{obs_table}. (\subref{ionization variability combined ph1}), (\subref{ionization variability combined ph2}) \&\ (\subref{ionization variability combined ph3}) Ionisation parameter ($\xi$) for the individual observations  derived using stacked spectrum. Compared to (\subref{light curve}) it seems $\xi$ varies in response to changing source luminosity. The response time is related to the gas density so the component in phase 3 is possibly denser than the remaining two warm-absorbers.}
    \label{ionization variability combined}
\end{figure*}

The spectrum used in our analysis compiled using 13 observations over a period of about 10 years. Using the best-fit parameters from the high resolution stacked spectrum we can now look at the source during the individual observations. For each spectrum we are primarily interested in the gas response to changes in source luminosity and so we keep all best-fit parameters from the stacked data frozen except for the normalisation of lines and continuum as well as the ionisation parameter ($\xi$). The resulting set of parameters is shown in Table 5 and all spectra are fitted with $\chi^{2} /d.o.f. \sim$1.2. Based on this in Fig. \ref{light curve} we plot the long term light-curve for Ark 564 where it is apparent that over this period the source seems to vary by a factor of 2. Unlike continuous monitoring surveys which allow analysis of evenly spaced time-intervals (e.g. \citet{2002A&A...391..875G}) here we have time-averaged observations taken over a decade. However, the 2011 data were taken every 6 days from May 2011 to July 2011 and so here we shall use this set to look for correlations. Additionally, we are interested in the gas response for which we use the $\xi$ parameter as a proxy. Fig. \ref{ionization variability combined}(b-d) shows the variation in ionisation over all observations where at first glance it seems that during 2011 all three gas phases seem to be responding to the ionising luminosity. However, within the shown errors only the highest ionised phase (\#\ 3) seems to mirror the light curves in Fig. \ref{light curve} unlike the other two phases of warm-absorbers.

\begin{figure}
\centering
\resizebox{6.5cm}{!}{\includegraphics[angle=0]{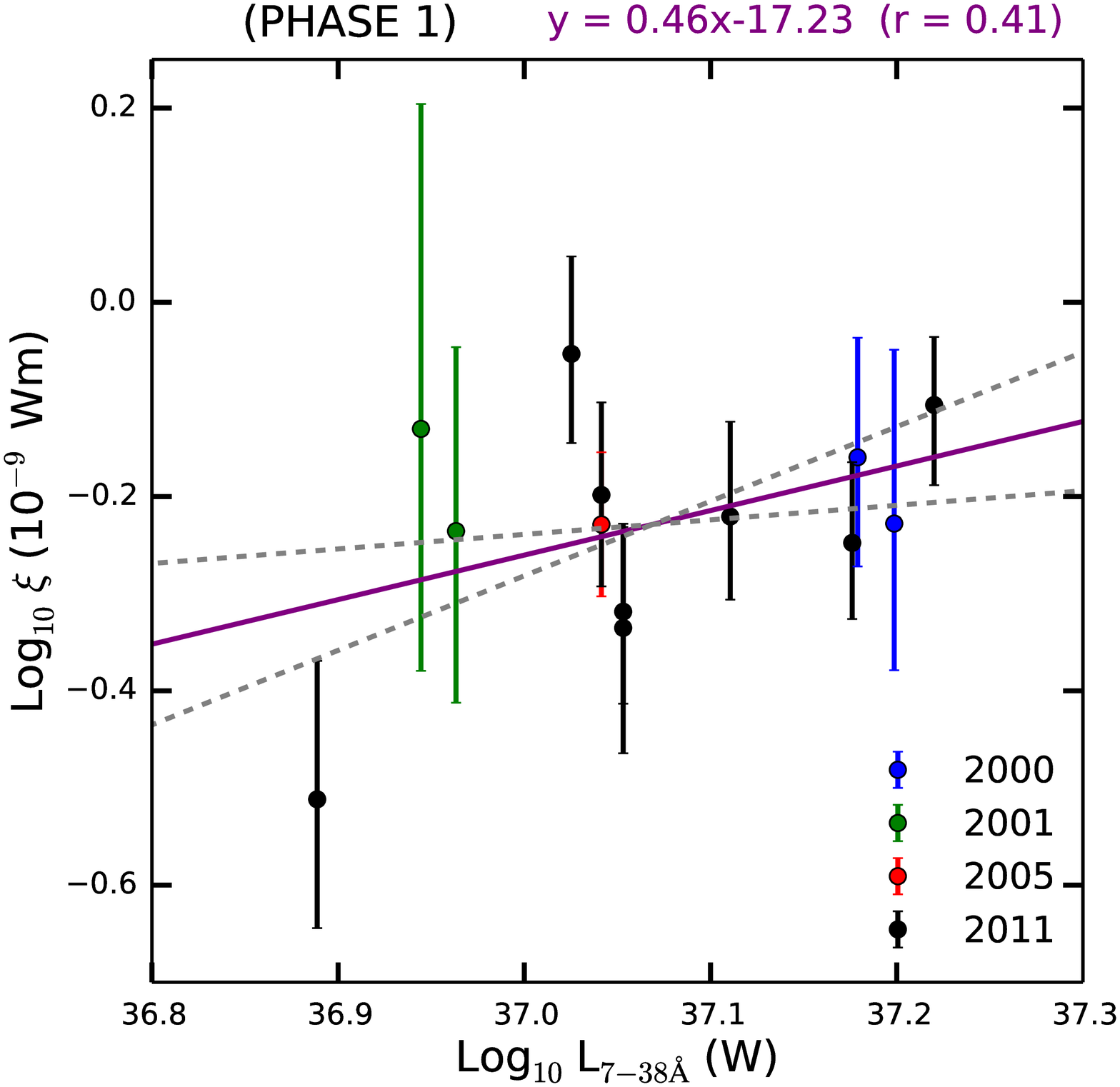}}
\resizebox{6.5cm}{!}{\includegraphics[angle=0]{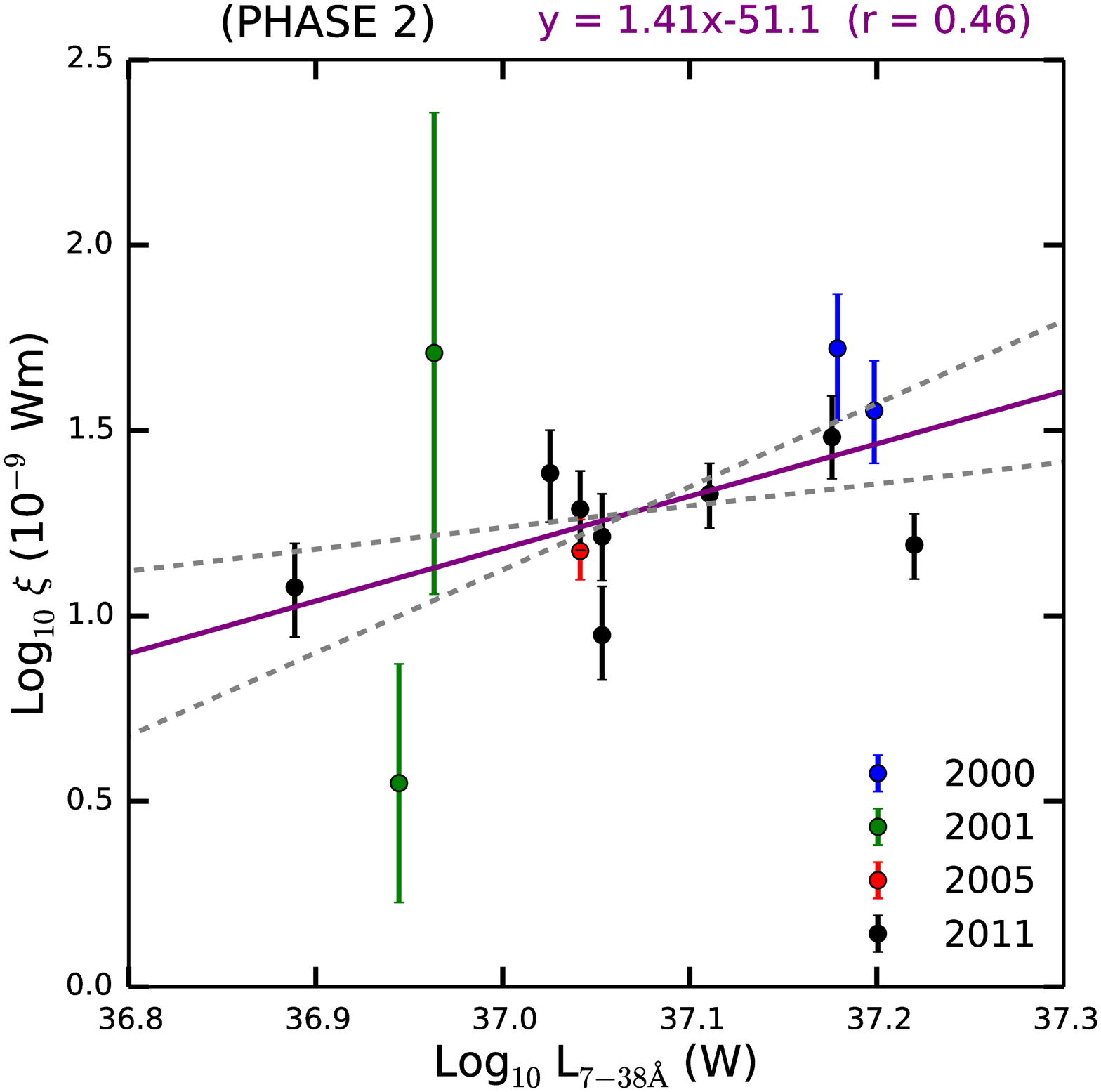}}
\resizebox{6.5cm}{!}{\includegraphics[angle=0]{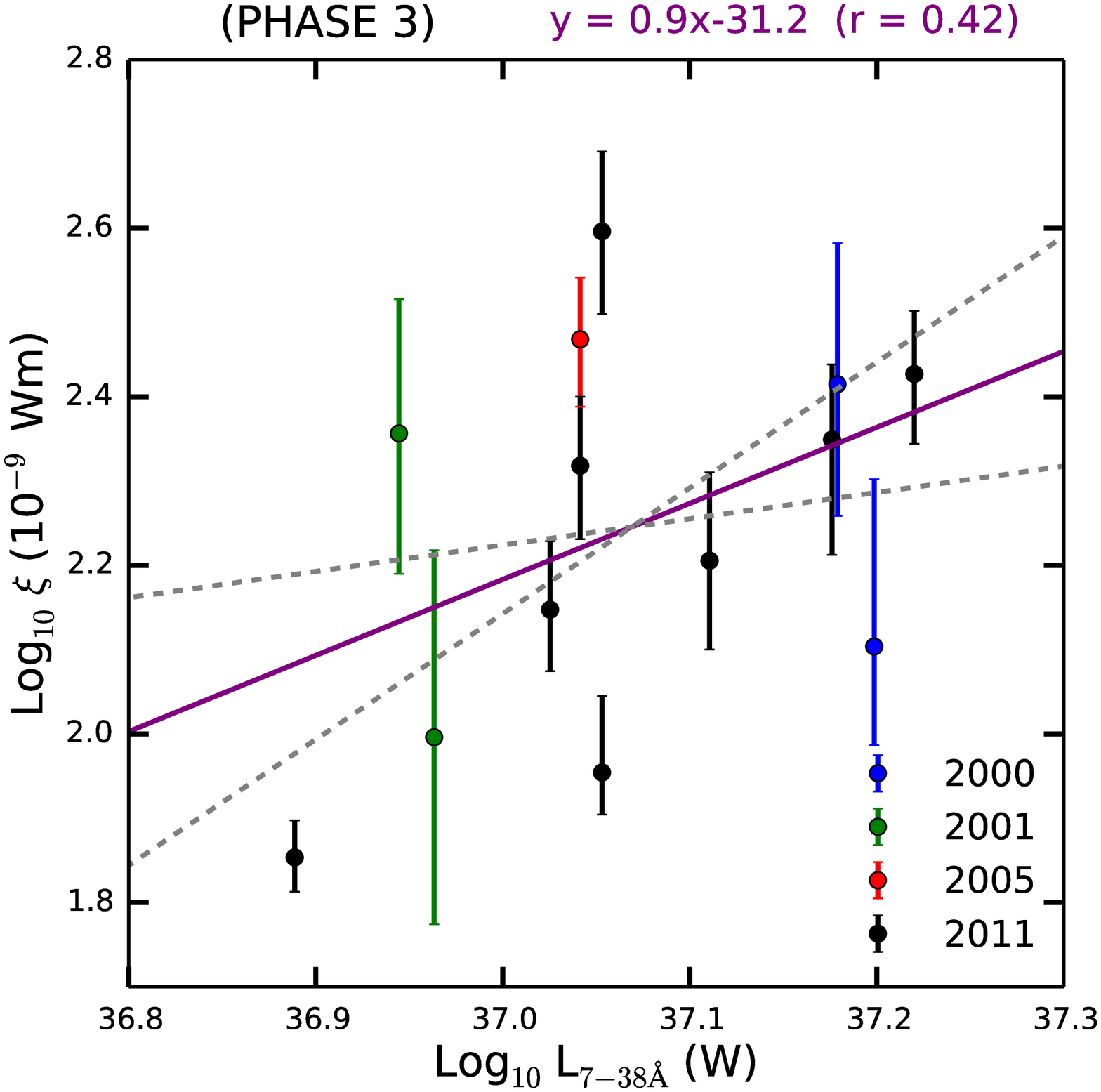}}
\caption{Ionisation plotted against Luminosity for all Ark 564 RGS observations. The panels correspond to the three phases and the best-fit line (purple) and Error in the best-fit (grey) along with the Pearson coefficient $r$ are also shown. Top: Phase 1 ($\xi$ $\simeq$-0.25, gradient: 0.46 $\pm$ 0.31); Centre: Phase 2 ($\xi$ $\simeq$1.37, gradient: 1.41 $\pm$ 0.82); Bottom: Phase 3 ($\xi$ $\simeq$2.38, gradient: 0.90 $\pm$ 0.59). It seems $\xi$ and $L$ are linked but in an imperfect correlation possibly due to sparse data for Ark 564.}
\label{xi_L plots}
\end{figure}

We can use recombination timescales to estimate the location of the warm-absorbers, the  clouds of photoionised gas being radiatively driven out. Essentially, the response timescale to luminosity variations can be used as a measure for the gas density as $\tau_{\mathrm{rec}} \sim \frac{1}{\alpha_{\mathrm{rec}} n_{\mathrm{e}}}$ such that denser gas would respond faster to changes in source luminosity \citep{1995MNRAS.273.1167R}. If the gas density is high enough then a rise in luminosity should be followed by a rise in $\xi$. This could be instantaneous (extremely dense) or there could be a significant lag (low density). In Fig. \ref{ionization variability combined}(b-d) we plot $\xi$ for the individual observations, derived by using the best-fit parameters from the stacked spectrum as described earlier. The three panels show the gas response in each phase over the 10 year period and visually it seems that the ionisation parameter was responding to the source luminosity during the 2011 campaign. To check this further in Fig. \ref{xi_L plots} we again plot the best-fit $\xi$ parameter but now against the ionising luminosity for all individual observations to search for correlations. We find that in all three phases there is a weak correlation with an average Pearson coefficient of $\sim$0.45 and p-value$\sim$10\%\ thus we can't reject null hypothesis. However, if in 2011 (evenly spaced data set) all phases change responding to source luminosity, this would indicate that the gas response time has to be less than or equal to $\sim$6 days (an assumption we use to estimate the distances to the warm-absorbers).
%
\subsection{Constraints on warm-absorber location}
\label{dist_est}
Since there are at least three independent absorbers, we can now estimate their locations, $R$, with respect to the source. By integrating the 2011 SED (Fig. \ref{ark564 SED 2011}), we can estimate the ionising luminosity, which between 1 and 1000 Rydberg gives L$_{\mathrm{ion}}$ = 6.9 $\times$ 10$^{37}$ W. This can be related to the absorbing source by the definition of $\xi$ (Sect. \ref{Absorption clouds}) and following \citet{2008A&A...490..103S}, we assume that the thickness ($\Delta r$) of any ionised gas phase has to be less than or equal to it's distance ($R$) from the source or:
\begin{equation}
\frac{\Delta r}{R} \leq 1
\end{equation}
and since the neutral Hydrogen column density ($N_{\mathrm{H}}$) can be related to the number density as:
\begin{equation}
N_{\mathrm{H}} \sim n(R)C_{\mathrm{v}} \Delta r
\end{equation}
we can provide an expression for the upper limit of the gas cloud location as: 
\begin{equation}
R \leq \frac{L_{\mathrm{ion}} C_{\mathrm{v}}}{\xi N_{\mathrm{H}}}
\label{upper-limit-equation}
\end{equation}
with the NLS1 volume filling factor, $C_{\mathrm{v}}$ estimated following \citet{2005A&A...431..111B} for each phase by matching outflow momentum with scattered and absorbed radiation momentum giving:
\begin{equation}
\label{eq_covering_factor}
C_{v} = \frac{(\dot{P}_{\mathrm{abs}} + \dot{P}_{\mathrm{scat}})\xi}{1.23 m_{p}L_{\mathrm{ion}}v^{2}\Omega}
\end{equation}
with the components defined in Appendix A and values listed in Table \ref{absorption phases}. Since phase 3 is consistent with negligible outflow we prefer not to average the remaining two covering factors but instead use the mean $C_{\mathrm{v}}$= 0.03 for NLS1s as obtained by \citet{2005A&A...431..111B}. Thus we obtain upper limits for the distances from the source: phase 1 = 74 kpc, phase 2 = 864 pc and phase 3 = 37 pc.

For a more robust estimate to the absorber distances we make use of the recombination time-scale, $\tau_{\mathrm{rec}}$. This defines how quickly the gas in the outflow is able to adjust to changes in ionisation and is related to the density $n_{\mathrm{H}}$ and recombination rate $\alpha_{\mathrm{r}}(X_{\mathrm{i}})$ from ion $X_{\mathrm{i+1}}$ to ion $X_{\mathrm{i}}$ by:
\begin{equation}		
\centering
      \tau_{\mathrm{rec}}(X_{\mathrm{i}}) = \left({\alpha_{\mathrm{r}}(X_{\mathrm{i}})n \left[\frac{f(X_{\mathrm{i+1}})}{f(X_{\mathrm{i}})} - \frac{\alpha_{\mathrm{r}}(X_{\mathrm{i-1}})}{\alpha_{\mathrm{r}}(X_{\mathrm{i}})}\right]}\right)^{-1}
\label{trec}
\end{equation}
where $f(X_{\mathrm{i}})$ is the fraction of all atoms of the element in ionisation state $X_{\mathrm{i}}$.

For each absorption phase, using the {\sc spex} tool `rec\_time' we calculate the product $n_{\mathrm{H}} \times t$ for the strongest lines (here Fe and O). Here, in principle, we should use ions with the highest concentrations but where $\tau_{\mathrm{rec}}$ is negative i.e. the ion recombines to a lower state we use the next strongest absorption after the first positive $\tau_{\mathrm{rec}}$ to avoid instability near turn-around in polarity. Then assuming that the fastest recombination would occur on the shortest time-scale of observations ($\tau_{\mathrm{obs}}\sim$ 6 days) we can obtain a lower limit for n$_{\mathrm{H}}$:
\begin{equation}
n_{H} \leq \frac{(n_{\mathrm{H}} \times t)}{\tau_{\mathrm{obs}}}
\end{equation}
and then using $ \xi = L/n r^{2}$, we obtain revised limits (Table \ref{absorption phases}) for the cloud locations with the innermost absorber being at about 4 pc from the source. \citet{2008A&A...490..103S} constrained the warm-absorbers to be between 100 pc$-$1000 kpc. We therefore report distance estimates improved by nearly 2 orders of magnitude. However, as was noted earlier, during the 2011 observations $\xi$ changes for all 3 phases (Fig. \ref{ionization variability combined}(b-d)) and the change is observed on the timescales of $\sim$6 days (the sampling period of the 2011 observations). But since the data are sparse it is possible the changes could be even faster; evidence of this is the correlation between $\xi$ and $L$ which is imperfect (Fig. \ref{xi_L plots}), the imperfections may point to the faster timescales. This suggests that the gas could be responding to luminosity changes in less than 6 days which would imply higher $n_{\mathrm{H}}$ and thus much lower distance estimates than those derived here.

\subsection{Outflow kinetics}

Using constraints on the location of the warm-absorbers we can now estimate the kinetics of the outflowing material. Following \citet{2012ApJ...753...75C} the mass outflow rate ($\dot{M}_{\mathrm{out}}$) and Kinetic Luminosity, $L_{\mathrm{kin}}$ are given by:
\begin{equation}
\dot{M}_{out} = 4\pi \mu m_{p} N_{H} C_{v} r v_{\mathrm{out}}
\end{equation}
\begin{equation}
L_{kin} = 2\pi \mu m_{p} N_{\mathrm{H}} C_{\mathrm{v}} r v_{\mathrm{out}}^{3}
\end{equation} where we use the outflow velocity (v$_{out}$) as in Table \ref{absorption phases}. Again to provide conservative estimates we use the typical average volume filling factor for NLS1's of $C_{\mathrm{v}}$= 0.03 similar to that obtained for phase 1 in Ark 564. Ignoring phase 3 (since it is consistent with negligible outflow), we find $\dot{M}_{\mathrm{out}}$ $\sim$ 0.1-2.0 $M_{\odot}$ yr$^{-1}$ and $L_{\mathrm{kin}}/L_{\mathrm{bol}} \sim$ 0.0002\%-0.01\% for phase 1 and phase 2 respectively. We note that in their analysis of 23 AGN    \citet{2012ApJ...753...75C} use a global covering factor of 0.5 for their estimates which would scale our numbers by two orders of magnitude. Given that the warm-absorbers in the case of Ark 564 are not co-located we prefer to use the average covering factor for this source. 
\subsection{Outflow strength and impact on host galaxy}
\label{outflow impact}

The outflow velocities for Ark 564 are lower than are generally expected of NLS1s,  given the high accretion rate relative to Eddington Luminosity \citep{2002ApJ...565...78B}. A number of factors could be responsible for the weak outflows. Firstly, if the outflows are radiation-pressure driven, the intrinsic variability of the source and thus of the accretion rate could naturally force material at different strengths.  Secondly, orientation of the source relative to the observer could play a role where the highest-velocity components will be detected if the outflowing gas is perfectly in the line of sight and so it follows that at any deviation from this angle only the transverse component would be seen. \citet{2013A&A...551A..95R} also find similar weak outflows in Ark 564 and suggest that this component could belong to the base of the wind in the accretion disk comparing it to the situation with NGC 4051 which is likely to be oriented at an angle of $\delta \sim$30$\degr$ \citep{2007ApJ...659.1022K}. In our analysis if at all the N VII and O VIII relativistic lines are statistically significant ($\Delta \chi^{2} \sim$ 500 for 1490 $d.o.f.$) we can thus support this argument as the {\sc laor} profile predicts a disk at inclination angle of 37$\degr \pm$0.5$\degr$.

Ark 564 does not seem to have extreme mass outflow rate and kinetic luminosity, with a maximum L$_{\mathrm{kin}}$/L$_{\mathrm{bol}}$ $\sim$ 0.01\%. Our estimates are very similar to those derived using the \chandra 2008 data analysed by \citet{2013ApJ...768..141G}. Moreover, \citet{2005A&A...431..111B} collated high-resolution spectroscopy results for 23 AGN and found the median $M_{\mathrm{acc}} \sim$0.04 $M_{\odot}$ yr$^{-1}$ and $\dot{M}_{out}$ $\sim$ 0.3 $M_{\odot}$ yr$^{-1}$ which would suggest that Ark 564 has kinetics of a very typical Seyfert. In order for AGN to impact their host galaxies, models require between L$_{\mathrm{kin}}$/L$_{\mathrm{bol}}$ $\sim$ 0.1\% \citep{2010MNRAS.401....7H} and $\sim$ 5\% \citep{2005Natur.433..604D} and so it would appear that in its current state there is no significant feedback from the AGN to its host environment in Ark 564. However, these ionised outflows are believed to be long-lived matching the typical AGN lifetime of $\sim$ 10$^{7}$ years during which the warm-absorbers could travel out to a few kpc and deposit up to 10$^{54}$ erg s$^{-1}$ even with weak outflows such as observed for Ark 564. This is only an order of magnitude smaller than typical energies required for heating the ISM to $T \sim$10$^{7}$K and affect star-formation \citep{2007ApJ...659.1022K}. Lastly, while the upper limit to the location of the phase 1 absorber (74 kpc) does suggest it might be sustained by processes different than the ones responsible for phase 2, the fact that both these absorbers have similar kinetics lends further support to the idea that ionised outflows could be long-lived and thus could affect the host galaxy over typical AGN lifetime.
\section{Summary}
\label{summary}
Ark 564 is a well studied source across different wavelengths and here we performed high-resolution X-ray spectroscopy on the combined data set from all \xmm observations of this source. We determined the SED using a broadband continuum model for AGN and carried out photoionisation modelling on it. We detect Gaussian-broadened emission lines from three distinct zones: C, N and O Ly$\alpha$ ($\sigma \sim$ 900 km s$^{-1}$), N VI intercombination (i) ($\sigma \sim$ 500 km s$^{-1}$) and O VII \&\ Ne IX forbidden (f) ($\sigma \sim$ 100 km s$^{-1}$). Significant presence of (i) and (f) relative to resonance lines also suggests photoionisation equilibrium (PIE) and so we used the \textit{xabs} model to calculate transmission through a slab of material ionised by the central source and detect three phases of warm-absorbers with different ionisation parameters (-0.25 to 2.38). Also, these gas phases are not in pressure equilibrium and are not co-located, with the highest ionised phase possibly beyond 4 pc from the source. We also report two possible broad relativistic lines, though the RGS calibration uncertainties around the same wavelengths cast doubt on their presence. Using results from the stacked spectrum we provide best-fit parameters for the individual observations and searched for variability of luminosity and gas response to this. We find that the ionisation parameter seems to follow changes in luminosity, although not in a fully correlated way. Though the outflow velocities found here were weak, we made estimates on the kinetics of outflowing gas and find that in its current state the AGN in Ark 564 is unlikely to disrupt the host galaxy although over the typical lifetimes of active galaxies enough energy can be deposited into the ISM such that star formation can be affected.
\begin{acknowledgements}

SRON is supported financially by NWO, the Netherlands Organization for Scientific Research.

\end{acknowledgements}


\newpage
\noindent{\bf APPENDIX A}
\label{covering-factor-appendix}
$\quad$\\
Following \citet{2005A&A...431..111B} we assume that the momentum of the outflow matches the sum of the momentum of the absorbed and scattered radiation. The basic equations used to estimate the volume filling factor are then as follows:
\begin{equation}
C_{v} = \frac{(\dot{P}_{\mathrm{abs}} + \dot{P}_{\mathrm{scat}})\xi}{1.23 m_{\mathrm{p}}L_{\mathrm{ion}}v^{2}\Omega}
\end{equation}
with the ionising luminosity $L_{\mathrm{ion}}$ = 6.9 $\times$ 10$^{37}$ W for Ark 564. The AGN solid angle is taken as $\Omega \sim$ 1.6. To calculate the absorbed luminosity we applied our \xabs parameters to the 2011 SED in order to obtain the transmission of each warm-absorber. 
Thus,
\begin{equation}
L_{\mathrm{abs}} = \frac{\Sigma Flux_{\mathrm{abs}} }{\Sigma Flux_{\mathrm{not-abs.}}} \times L_{\mathrm{ion}}
\end{equation}
where the summation is over all energy bins. This gives the expression for absorbed momentum as:
\begin{equation}
\dot{P}_{\mathrm{abs}} = \frac{L_{\mathrm{abs}}}{c}
\end{equation}
For the scattered radiation momentum we use the optical depth ($\tau _{\mathrm{T}} = \sigma _{T} N_{\mathrm{H}}$) of the absorber with column density $N_{\mathrm{H}}$ and Thomson cross-section $\sigma _{T}$ to obtain:
\begin{equation}
\dot{P}_{\mathrm{scat}} = \frac{L_{\mathrm{ion}}}{c} (1- e^{-\tau_{T}})
\end{equation}

Table \ref{absorption phases} lists the $L_{\mathrm{abs}}$ for phases 1 and 2 which both show outflows. Note that in this scheme phase 3 is closest to the source and so phase 1 is the outermost absorber. Also, since Phase 3 does not show any outflow we can not estimate a physical volume filling factor for this particular absorber unlike the other two mentioned above.
\section*{\bf APPENDIX B}
\label{Best-fit for individual observations}
Table 5 on the following page presents the best-fit parameters obtained for the individual observations as discussed in Sect. \ref{variab_results}. This was achieved by freezing all best-fit parameters from the stacked spectrum except for the flux normalisation and ionisation parameter ($\xi$).
%
\newcounter{tempfootnote}
\setcounter{tempfootnote}{\value{footnote}}
\setcounter{footnote}{0}
\renewcommand{\thefootnote}{\alph{footnote}}
\begin{sidewaystable*}
\begin{center}
\begin{minipage}[t]{22cm}
\setlength{\extrarowheight}{22pt}
\caption{Best fit for individual spectra from 2000-2011 as discussed in Sect. \ref{variab_results}. All spectra are fitted with $\chi^{2} /d.o.f. \sim$1.2}
\renewcommand{\footnoterule}{}
\resizebox{\textwidth}{!}{\begin{tabular}{l|l|l|l|l|l|l|l|l|l|l|l|l|l}
\hline
Param & \#\ 01 & \#\ 02 & \#\ 03 & \#\ 04 & \#\ 05 & \#\ 06 & \#\ 07 & \#\ 08 & \#\ 09 & \#\ 10 & \#\ 11 & \#\ 12 & \#\ 13 \\
\hline
Log $\xi$\footnote{Ionisation parameter in 10$^{-9}$ Wm.} (Phase 1) & -0.23 $\pm$ 0.13 & -0.16 $\pm$ 0.10 & -0.24 $\pm$ 0.20 & -0.13$^{+0.34} _{-0.23}$ & -0.25 $\pm$ 0.07 & -0.16 $\pm$ 0.10 & -0.10 $\pm$ 0.09  & -0.18 $\pm$ 0.09 & -0.26 $\pm$ 0.08 & -0.35 $\pm$ 0.09 & -0.50 $\pm$ 0.15 & -0.34 $\pm$ 0.12 & -0.25 $\pm$ 0.10\\

Log $\xi$ $^{\mathrm{a}}$ (Phase 2) & 1.56 $\pm$ 0.10 & 1.73 $\pm$ 0.13 &  1.71 $\pm$ 0.65 & 0.55 $\pm$  0.29 & 1.25 $\pm$ 0.07 & 1.21 $\pm$ 0.09 & 1.41 $\pm$ 0.10  & 1.36 $\pm$ 0.09 & 1.34 $\pm$ 0.07 & 1.24 $\pm$ 0.10 & 1.07 $\pm$ 0.13 & 0.95 $\pm$ 0.12 &  1.50 $\pm$ 0.10 \\

Log $\xi$ $^{\mathrm{a}}$ (Phase 3) & 2.07 $\pm$ 0.12 &  2.41 $\pm$ 0.20 & 1.99 $\pm$ 0.20 & 2.36 $\pm$ 0.15 & 2.41 $\pm$ 0.08 & 2.43 $\pm$ 0.08 & 2.12 $\pm$ 0.07 &  2.29 $\pm$ 0.08 & 2.22 $\pm$ 0.11 & 1.95 $\pm$ 0.06 & 1.85 $\pm$ 0.05 &  2.60 $\pm$ 0.10 &   2.35 $\pm$ 0.10 \\

O VII (f)\footnote{Non-relativistic emission normalisation in 10$^{50}$ ph s$^{-1}$.}  & 0.3$^{+0.7} _{-0.3}$ &  2.4 $\pm$ 0.6 & 1.1 $\pm$ 0.9 & 0.8 $\pm$ 0.6 & 0.9 $\pm$ 0.3 & 0.6 $\pm$ 0.4 & 1.3 $\pm$  0.4 & 0.4 $\pm$ 0.3 & 7.7 $\pm$ 0.3 & 1.0 $\pm$ 0.4 & 1.0 $\pm$ 0.3 &  1.1 $\pm$ 0.4 & 0.9 $\pm$ 0.4\\

Ne IX (f) $^{\mathrm{b}}$  & 0.8 $\pm$ 0.3 &  0.3 $\pm$ 0.2 &  0.8 $\pm$ 0.4 &  0.6 $\pm$ 0.2 & 0.3 $\pm$ 0.1 & 0.4 $\pm$ 0.2 & 0.5 $\pm$ 0.1 & 0.2 $\pm$ 0.1 & 0.3 $\pm$ 0.1 & 0.3 $\pm$ 0.1 & 0.4 $\pm$ 0.1 & 0.3 $\pm$ 0.2 &  0.2 $\pm$ 0.2\\

N VII Ly$\alpha$ $^{\mathrm{b}}$  & 0.3 $^{+1.1} _{-0.3}$ & 1.6 $\pm$ 1.1 & 3.6 $\pm$ 1.7 & 2.8 $\pm$ 1.4 & 2.4 $\pm$ 0.4 & 2.6 $\pm$ 0.6 & 2.3 $\pm$ 0.5 & 2.4 $\pm$ 0.5 & 2.8 $\pm$ 0.5 &  2.2 $\pm$ 0.5 & 2.9 $\pm$ 0.7 & 2.9 $\pm$ 7.4 & 3.5 $\pm$ 0.9\\ 

N VI (i) $^{\mathrm{b}}$ & 0.2$^{+1.0} _{-0.2}$ &  2.6 $\pm$ 0.8 & 1.7 $\pm$ 1.1 & 1.5 $\pm$ 0.8 & 1.3 $\pm$ 0.4 & 1.2 $\pm$ 0.6 & 1.1 $\pm$ 0.5 & 1.4 $\pm$ 0.5 & 1.2 $\pm$ 0.5 & 0.8 $\pm$ 0.5 & 0.9 $\pm$ 0.4 & 0.9 $\pm$ 0.5 &  1.3 $\pm$ 0.6\\  

O VIII Ly$\alpha$ $^{\mathrm{b}}$ & 1.8 $\pm$ 0.7 & 2.8 $\pm$ 0.7 &  1.7 $\pm$ 1.0 & 2.7 $\pm$ 0.7 & 1.4 $\pm$ 0.3 & 1.3 $\pm$ 0.4 & 1.7 $\pm$ 0.3 & 1.8 $\pm$ 0.3 & 2.1 $\pm$ 0.3 & 2.0 $\pm$ 0.3 & 1.5 $\pm$ 0.3 & 1.3 $\pm$ 0.4 & 2.1 $\pm$ 0.4\\

C VI Ly$\alpha$ $^{\mathrm{b}}$ & 2.2 $\pm$ 1.7 & 0$^{+1.3}$  & 0$^{+1.5}$ &  2.2 $\pm$ 2.2 & 0.4 $^{+0.8} _{-0.4}$ & 0.1$^{+1.0}$ & 2.7 $\pm$ 1.1 & 1.9 $\pm$ 1.1 & 43.0 $^{+0.6} _{-43.0}$ & 1.3 $\pm$ 1.1 & 1.2 $\pm$ 1.2 & 1.1 $\pm$ 1.1 & 3.1 $\pm$ 1.5\\

O laor\footnote{ Broad relativistic line normalisation in 10$^{50}$ ph s$^{-1}$.} & 12 $\pm$ 5 & 10 $\pm$ 4 & 20 $\pm$ 6 & 2 $^{+3}$ & 3 $\pm$ 2 & 2 $^{+3}$ & 8 $\pm$ 2 & 7 $\pm$ 2 & 6 $\pm$ 2 & 1 $^{+2}$ & 5 $\pm$ 2 & 8 $\pm$ 2 & 12 $\pm$ 3\\

N laor $^{\mathrm{c}}$ & 1 $\pm$ 1 & 0$^{+16}$ & 35 $\pm$ 21 &  15 $\pm$ 15 & 12 $\pm$ 6 & 18 $\pm$ 9 & 10 $\pm$ 8 & 14 $\pm$ 8 & 20 $\pm$ 8 & 8 $\pm$ 8 & 12 $\pm$ 8 & 30 $^{+10} _{-1}$ & 36 $\pm$ 10\\
\hline
\end{tabular}}
\end{minipage}
\end{center}
\end{sidewaystable*}
\setcounter{footnote}{\value{tempfootnote}}
\renewcommand{\thefootnote}{\arabic{footnote}}

\end{document}